\documentclass[thmsa,12pt]{report}
\usepackage{sw20jrep}

\input tcilatex
\QQQ{Language}{
American English
}

\begin{document}

\author{Frank B. Estabrook \and R. Steve Robinson \and Hugo D. Wahlquist \\
Jet Propulsion Laboratory 169-327\\
California Institute of Technology\\
4800 Oak Grove Drive\\
Pasadena, CA 91109, USA}
\date{October 31, 1996}
\title{Hyperbolic equations for vacuum gravity using special orthonormal frames}
\maketitle

\begin{abstract}
By adopting Nester's $4-$dimensional special orthonormal frames, the tetrad
equations for vacuum gravity are put into explicitly causal and symmetric
hyperbolic form, independent of any time slicing or other gauge or
coordinate specialization.
\end{abstract}

\section{Introduction}

We have previously given a well set and causal exterior differential system
for vacuum gravity, generated by a closed set of differential forms
describing the immersion of $4-$dimensional spacetime into a flat $10$
dimensional space \cite{E&W93}. The orthonormal frame bundle of the latter
has a canonical basis of $10$ (translation) $1-$forms $\omega ^\mu $ and $45$
(rotation) $1-$forms $\omega _\nu ^\mu $, satisfying the structure equations
of the Lie group $ISO(10)$. Dividing the range $\mu ,\nu =1,\cdots ,10$ into
two ranges $i,j=1,\cdots ,4$ and $A,B=5,\cdots ,10$, these are 
\[
d\omega ^i+\omega _j^i\wedge \omega ^j+\omega _B^i\wedge \omega ^B=0 
\]
\[
d\omega ^A+\omega _j^A\wedge \omega ^j+\omega _B^A\wedge \omega ^B=0 
\]
\[
d\omega _k^i+\omega _j^i\wedge \omega _k^j+\omega _B^i\wedge \omega _k^B=0 
\]
\[
d\omega _A^i+\omega _j^i\wedge \omega _A^j+\omega _B^i\wedge \omega _A^B=0 
\]
\[
d\omega _C^A+\omega _j^A\wedge \omega _C^j+\omega _B^A\wedge \omega _C^B=0. 
\]

The exterior differential system is generated by $6$ (immersion) $1-$forms $%
\omega ^A$, their closure $2-$forms $d\omega ^A=-\omega _i^A\wedge \omega ^i$%
, and $4$ closed $3-$forms ensuring Ricci-flatness, namely, $R_{ij}\wedge
\omega _k\,\varepsilon ^{ijkl}$, where $\frac 12R_j^i=d\omega _j^i+\omega
_k^i\wedge \omega _j^k=-\omega _A^i\wedge \omega _j^A$ defines the Riemann $%
2-$forms.

The significant result is the calculation of the Cartan characteristic
integers 
\[
s=\{s_0,s_1,s_2,s_3\}=\{6,6,10,8\}. 
\]
This shows the solutions to be $25$ dimensional, regular (i.e., in principle
constructed from a series of Cauchy-Kowalewsky integrations) and causal
(i.e., $s_4=0$, so the solutions are determined from suitable data set on $3$
dimensional immersed manifolds). The solutions are involutory with respect
to $\omega ^i,\omega _j^i$ and $\omega _B^A$: these forms remain independent
when pulled back into a solution manifold and can be adopted as a basis
there.

The six basis forms $\omega _j^i$ and $15$ basis forms $\omega _B^A,$ which
occur of course in the structure relations, do not appear explicitly in the
exterior differential system, showing the solutions to be bundles having $21$
dimensional fibers over a four dimensional base. Evidently this expresses
arbitrary $O(4)$ (for Lorentzian, $O(3,1)$) rotations of the tetrad frame $%
\omega ^i$ and $O(6)$ rotations of the immersion co-frame $\omega ^A$ at
each point of the base. On a four dimensional cross-section all the forms
can be expanded on the $\omega ^i$ basis, the $\omega _j^i$ being a metric
connection.

In Section 2 we give a new immersion exterior differential system for vacuum
gravity that incorporates the higher dimensional special orthonormal frame
(HSOF) conditions proposed by Nester \cite{N92} as a generalization of the
special orthonormal frame conditions (SOF) in three dimensional Riemannian
geometry \cite{N91} \cite{N89} \cite{D&M}. \textbf{Calculation of the Cartan
characteristic integers for this exterior differential system for }$4$%
\textbf{\ dimensional Riemannian geometry shows that it is also well set and
causal, so that, as Nester conjectured, such frames can be imposed without
impediment in extended regions of vacuum spacetime.} Solutions are now $19$
dimensional, the fibers expressing only arbitrary $O(6)$ rotations of the
co-frames.

These exterior differential systems include all integrability conditions for
the determination of a metric on the base, or on any $4$ dimensional cross
section. In terms of base space coordinates $x^\mu $ $(\mu ,\nu =1,2,3,4)$,
invertible matrices of functions $^i\lambda _\mu (x)$ and $_i\lambda ^\mu (x)
$ exist such that $^i\lambda _{\mu \;j}\lambda ^\mu =\delta _j^i,$ $%
(i,j=1,2,3,4).$ We introduce the Minkowski metric, i.e., $\eta ^{i\,j}=%
\mathtt{diag}(1,1,1,-1)$, to raise and lower left, or Lorentz, indices. Then
the metric is given by $g^{\mu \,\nu }=\eta ^{k\,l}\;_k\lambda ^\mu
\;_l\lambda ^\nu .$ Inserting $\omega ^i=\;^i\lambda _\mu \;dx^\mu $ in the
structure equations and in the exterior differential system gives the
partial differential equations for coordinate components of the tetrad field.

In Section 3 we reformulate the higher dimensional special orthonormal frame
system in orthonormal tetrad components, using $\eta ^{i\,j}$. We use the
dyadic formalism of references \cite{E&W65} and \cite{Wahl}, in which the
unit $4-$vector field $^4\lambda ^\mu $ is given a special meaning: it
traces a congruence of timelike world lines, a $3$ parameter ``fluid'' of
point observers, to which physical interpretations of the $24$ components of
the connection are attributed. The three spacelike unit vectors $^a\lambda
^\mu \;(a=1,2,3)$ at each point complete a local orthonormal frame for the
observer there. We expand the six $1-$forms $\omega _j^i$ on the $\omega
^a\!,\omega ^4$ basis ($a,b=1,2,3$), and we expand the six $2-$forms $R_j^i$
on the $\omega ^a\wedge \omega ^b\!,\omega ^a\wedge \omega ^4$ basis,
subject to the conditions for Ricci-flatness.

The 24 connection components are grouped into the following ensembles: $%
3\times 3$ dyadics $\mathsf{K}$ and $\mathsf{N}$,\textsf{\ }and\textsf{\ }$%
3- $vectors $\mathbf{a}$ and $\mathsf{\omega }$. The dyadic $\mathsf{K}$ has
components $K_{ab}$, and can be resolved into 
\[
K_{ab}=S_{ab}-\Omega _c\,\varepsilon _{acb}\text{,} 
\]
where $S_{ab}$ is the symmetric rate-of-strain $3-$tensor of the observer
fluid, and $\Omega _c$ is its axial vector of vorticity. Or, in dyadic
notation 
\[
\mathsf{K=S-\,}\mathbf{\Omega }\mathsf{\times I}, 
\]
where \textsf{I} is the unit dyadic. The dyadic $\mathsf{N}$ is formed from
the nine spacelike Ricci rotation coefficients of the $\omega ^a$ basis. It
has components $N_{ab}$ and can similarly be resolved into a symmetric part
and an axial vector $n_c$ 
\[
N_{ab}=N_{ab}^{sym}-n_c\,\varepsilon _{acb}\text{.} 
\]
Again we write in dyadic notation, 
\[
\mathsf{N=N}^{sym}-\mathbf{n}\times \mathsf{I.} 
\]
The vector $\mathsf{\omega }$ has components $\omega _a$ and is the
time-dependent angular velocity of the triad seen by an observer moving
along $\omega ^4$, with respect to a Fermi-propagated frame. Since $\mathsf{%
\omega }$\textsf{\ }is a standard notation for angular velocity this usage
should in context not be difficult to distinguish from the $1-$forms $\omega
^a$ we have used up to this point. (When necessary, we can write the
components of the angular velocity as $\overline{\omega }_a.$) The vector $%
\mathbf{a}$ has components $a_a$ and is the acceleration of the point
observers, i.e., their departure from geodesic motion. The vectors $\mathbf{a%
}$ and $\mathsf{\omega }$ are in principle determined operationally by the
point observers, using spring balances and supported spinning particles in
the local frames. The ten components of the Weyl tensor yield symmetric
tracefree dyadics \textsf{A} and \textsf{B}, the ``electric'' and
``magnetic'' tidal fields. The quantities \textsf{A}, \textsf{B}, $\mathsf{K}
$, \textsf{N},\textbf{\ a},\textbf{\ }and $\mathsf{\omega }$\textsf{\ }are
defined in terms of the $_i\lambda _{\mu ,\nu }$ in \cite{E&W65}.

The dyadic formalism is completed by the use of inner and outer $3-$%
dimensional multiplication ( $\cdot $ and $\times $ ), and by convective
derivatives (also known as unit derivations) in the timelike and spacelike
directions ( $\mathbf{\dot{ }}$ and $\mathbf{D}$ ). Use of spatial covariant
differentiation $\mathsf{\nabla }$, related to $\mathbf{D}$ by the spatial
connection $\mathsf{N}$, often is more efficient than the use of $\mathbf{D}$%
. Detailed expositions of this formalism, including the various $3-$vector
and dyadic relations, the relations of $\mathbf{D}$ and $\mathsf{\nabla }$
(involving $\mathsf{N}$),\textsf{\ }and the 52 general first order dyadic
differential equations for vacuum $4-$geometry, are to be found in
references \cite{E&W65} and \cite{Wahl}.

There are some notational changes of which the reader should be aware. The
dyadic we now (and in \cite{Wahl}) denote by $\mathsf{N}$\textsf{\ }was in 
\cite{E&W65} denoted by $\mathsf{N}^{\star }$. For the dyadic $\mathsf{N}$
in \cite{E&W65} (which was symmetric) one should now understand 
\[
\mathsf{N}-(Tr\,\mathsf{N)\,I\;+\;}\mathbf{n}\times \mathsf{I} 
\]
and for the vector $\mathbf{L}$ in \cite{E&W65} one should now write $%
\mathbf{n}.$

Section 3 also gives the 12 new equations specializing to $4-$dimensional
special orthonormal frames. Six of the additional equations involve ${%
\mathbf{\dot{a}}}$ and $\mathbf{\dot{\QTR{mathsf}{\omega}}}$, which had not
otherwise appeared in the tetrad equations. This is also the case with the
six conditions for the ``Lorentzian frame'' specialization (based on time
slicing) discussed by van Putten and Eardley \cite{vP&E}. When time
evolution of all dependent variables is explicit the causal structure of the
system can be seen as resulting from local wave propagation subject to
constraints expressing integrability conditions. The integrability
properties of these sets of dyadic equations are briefly discussed in
Section 3 in terms of Cartan's reduced characters and Cartan's test.

In Section 4 we present the complete set of 64 dyadic equations resulting
from substitution of dyadic components in the structure equations and the
HSOF exterior differential system of Section 2. There are 34 equations
involving the timelike derivatives of the 24 connection coefficients and the
10 Weyl components, together with an additional 30 transverse equations in
which no time derivatives appear. By forming appropriate linear
combinations, the final equations have been arranged to show that they have
first order symmetric hyperbolic (FOSH) structure similar to that of recent
formalisms involving preferred time slices, designed for application to
numerical gravity \cite{C&Y} \cite{AA&C} \cite{AAC&Y96} \cite{B&M89} \cite
{B&M92}.

A number of known and new ``hyperbolic reductions'' of the vacuum field
equations, and the choices of gauge they allow, have been surveyed by
Friedrich \cite{Fried96}. He uses both tetrad formalism, and ADM variables
based on preferred time slicing. The r\^{o}le of the Bianchi identities is
emphasized: the equations for \textsf{A} (in \cite{Fried96}, \textsf{E}) and 
\textsf{B}\textbf{,} which propagate as spin$-2$ massless fields (cf., e.g., 
\cite{E&W65}), are given in FOSH form. The FOSH equations obtained in the
present paper necessarily include these. Nester's conditions on the tetrad
components of the connection however leave no further freedom in choice of
gauge (or frame), and seem not to have been previously used. \textbf{They
result in FOSH structure with constant coefficients, and force all the
dependent variables to evolve along null cones.}

An analysis using Cartan's test shows that in the present case the observer
fluid necessarily has vorticity, i.e., $\mathbf{\Omega }\neq 0$, so it is
not $3-$space orthogonal, and therefore Nester's special orthonormal frames
cannot be based on a preferred slicing. Nevertheless, the relative
simplicity of these equations may be advantageous. No lapse or shift
variables or higher order derivatives have been introduced. Of course,
useful related coordinates can still be found. Two are suggested immediately
by the closed $2-$forms in the exterior differential system, and we note in
Appendix $1$ that the dyadic conditions for a harmonic timelike coordinate
are again of first order symmetric hyperbolic (FOSH) form, so can simply be
added to our results. Further adopting three spacelike coordinates co-moving
with the point observers then allows an explicit line element to be written,
given a solution of the FOSH\ equations.

A similar application of special orthonormal frames can be made for $2+1$
gravity, again leading to a constant coefficient FOSH system. This is
outlined in Appendix 2.

\section{Higher dimensional special orthonormal frames}

Using eight new variables $y_i$ and $z_i\,(i=1,2,3,4)$, we prolong the
previously given immersion exterior differential system for vacuum gravity,
with two additional closed $2-$forms, two additional $3-$forms, and their
closure $4-$forms. The $3-$forms and $4-$forms essentially define the $y_i$
and $z_i$ in terms of the connection forms (the $\omega _j^i$ pulled back
into the solutions), and the $2-$forms require them to satisfy Dirac type
partial differential equations. When expanded in tetrad components in
Section 3 it can be verified that this is precisely Nester's prescription.
The generators of this exterior differential system in 63 dimensions are 
\[
\omega ^A 
\]
\[
\omega _i^A\wedge \omega ^i 
\]
\[
(dy_i-\omega _i^jy_j)\wedge \omega ^i 
\]
\[
(dz_i-\omega _i^jz_j)\wedge \omega ^i 
\]
\[
(\omega _i^j\wedge \omega ^k\wedge \omega ^l+\tfrac 23y_i\,\omega ^j\wedge
\omega ^k\wedge \omega ^l)\varepsilon _{\cdot jkl}^i 
\]
\[
\omega _{ij}\wedge \omega ^i\wedge \omega ^j-\tfrac 13z_i\omega ^j\wedge
\omega ^k\wedge \omega ^l\varepsilon _{\cdot jkl}^i 
\]
\[
\omega _j^A\wedge \omega _k^A\wedge \omega ^i\varepsilon _i^{\cdot jkl} 
\]
\[
(-\omega _s^j\wedge \omega _i^s\wedge \omega ^k\wedge \omega ^l+2\omega
_i^j\wedge \omega _s^k\wedge \omega ^s\wedge \omega ^l+\tfrac 23dy_i\wedge
\omega ^j\wedge \omega ^k\wedge \omega ^l-2y_i\omega _s^j\wedge \omega
^s\wedge \omega ^k\wedge \omega ^l)\varepsilon _{\cdot jkl}^i 
\]
\begin{equation}
-\omega _{is}\wedge \omega _j^s\wedge \omega ^i\wedge \omega ^j+2\omega
_{ij}\wedge \omega _s^i\wedge \omega ^s\wedge \omega ^j-(\tfrac 13dz_i\wedge
\omega ^j\wedge \omega ^k\wedge \omega ^l-z_i\omega _s^j\wedge \omega
^s\wedge \omega ^k\wedge \omega ^l)\varepsilon _{\cdot jkl}^i.  \tag{1}
\end{equation}
Monte Carlo calculations \cite{E&W93} of the Cartan characteristic integers
yield 
\[
s=\left\{ 6,8,14,16\right\} , 
\]
therefore the exterior differential system is well set and causal; solutions
are 19 dimensional, fibered over 4 dimensions. The $\omega _B^A$ do not
appear in the exterior differential system, so the fibers express co-frame
rotation. But the $\omega _j^i$ are explicitly present so the frames are
specialized and determined up to a simultaneous rotation at every point.

If the two $2-$forms are exact, two further variables can be added, say $%
\zeta $ and $\eta $, together with $1-$forms 
\[
d\zeta -y_i\omega ^i, 
\]
\[
d\eta -z_i\omega ^i. 
\]
These should be useful for introducing intrinsic coordinates into the HSOF
formulation.

\section{The dyadic components of the connection forms $\omega _j^i$ and
Riemann forms $R_j^i$}

We expand the $\omega _j^i$ on $\omega ^i$, defining the 24 $3-$dyadic and $%
3-$vector components of $\mathsf{K}$, $\mathsf{N}$, $\mathsf{\omega }$, and $%
\mathbf{a}\mathsf{,}$ summarized in the Introduction and described in detail
in references \cite{E&W65} and \cite{Wahl} 
\[
\omega _{12}=-\omega _{21}=N_{13}\omega ^1+N_{23}\omega ^2+N_{33}\omega ^3+%
\overline{\omega }^3\omega ^4 
\]
\[
\omega _{23}=-\omega _{32}=N_{21}\omega ^2+N_{31}\omega ^3+N_{11}\omega ^1+%
\overline{\omega }^1\omega ^4 
\]
\[
\omega _{31}=-\omega _{13}=N_{32}\omega ^3+N_{12}\omega ^1+N_{22}\omega ^2+%
\overline{\omega }^2\omega ^4 
\]
\[
\omega _{14}=-\omega _{41}=-K_{11}\omega ^1-K_{21}\omega ^2-K_{31}\omega
^3+a^1\omega ^4 
\]
\[
\omega _{24}=-\omega _{42}=-K_{22}\omega ^2-K_{32}\omega ^3-K_{12}\omega
^1+a^2\omega ^4 
\]
\begin{equation}
\omega _{34}=-\omega _{43}=-K_{33}\omega ^3-K_{13}\omega ^1-K_{23}\omega
^2+a^3\omega ^4.  \tag{2}
\end{equation}
We also expand the Riemann $2-$forms on a basis consisting of $\omega
^a\wedge \omega ^4$ and $\omega ^a\wedge \omega ^b\,(a,b=1,2,3)$, to define
(in the Ricci-flat case) Weyl dyadics $\mathsf{A}$ and $\mathsf{B}$ 
\[
\tfrac 12R_{21}=B_{31}\omega ^1\wedge \omega ^4+B_{32}\omega ^2\wedge \omega
^4+B_{33}\omega ^3\wedge \omega ^4+A_{31}\omega ^2\wedge \omega
^3+A_{32}\omega ^3\wedge \omega ^1+A_{33}\omega ^1\wedge \omega ^2 
\]
\[
\tfrac 12R_{32}=B_{12}\omega ^2\wedge \omega ^4+B_{13}\omega ^3\wedge \omega
^4+B_{11}\omega ^1\wedge \omega ^4+A_{12}\omega ^3\wedge \omega
^1+A_{13}\omega ^1\wedge \omega ^2+A_{11}\omega ^2\wedge \omega ^3 
\]
\[
\tfrac 12R_{13}=B_{23}\omega ^3\wedge \omega ^4+B_{21}\omega ^1\wedge \omega
^4+B_{22}\omega ^2\wedge \omega ^4+A_{23}\omega ^1\wedge \omega
^2+A_{21}\omega ^2\wedge \omega ^3+A_{22}\omega ^3\wedge \omega ^1 
\]
\[
\tfrac 12R_{41}=-A_{11}\omega ^1\wedge \omega ^4-A_{12}\omega ^2\wedge
\omega ^4-A_{13}\omega ^3\wedge \omega ^4+B_{11}\omega ^2\wedge \omega
^3+B_{12}\omega ^3\wedge \omega ^1+B_{13}\omega ^1\wedge \omega ^2 
\]
\[
\tfrac 12R_{42}=-A_{22}\omega ^2\wedge \omega ^4-A_{23}\omega ^3\wedge
\omega ^4-A_{21}\omega ^1\wedge \omega ^4+B_{22}\omega ^3\wedge \omega
^1+B_{23}\omega ^1\wedge \omega ^2+B_{21}\omega ^2\wedge \omega ^3 
\]
\[
\tfrac 12R_{43}=-A_{33}\omega ^3\wedge \omega ^4-A_{31}\omega ^1\wedge
\omega ^4-A_{32}\omega ^2\wedge \omega ^4+B_{33}\omega ^1\wedge \omega
^2+B_{31}\omega ^2\wedge \omega ^3+B_{32}\omega ^3\wedge \omega ^1.\text{ }%
(3) 
\]
The Riemann $2-$forms given above are antisymmetric under interchange of
their indices. The 10 Weyl components satisfy $A_{ab}=A_{ba},$ $A_{aa}=0,$ $%
B_{ab}=B_{ba},$ and $B_{aa}=0.$ It can be verified that the Riemann
symmetries are satisfied, namely, 
\[
R_b^a\wedge \omega ^b+R_4^a\wedge \omega ^4=0 
\]
and 
\[
R_b^4\wedge \omega ^b=0, 
\]
and also that the four $3-$form conditions for Ricci-flatness, namely, 
\[
R_j^i\wedge \omega ^k\,\varepsilon _{\cdot ikl}^j=0 
\]
have been imposed.

From the $3-$forms of (1) we can now find the component expansions of the
fields $y_i$ and $z_i$, on an integral manifold of the exterior differential
system. These are given by the following 
\begin{equation}
y_i=(y^a,-y^4)=(\mathsf{N\dot{\times}I-}\mathbf{a}\mathsf{,+}Tr\mathsf{\,K})
\tag{4}
\end{equation}
\begin{equation}
z_i=(z^a,-z^4)=(\mathsf{-K\dot{\times}I+\omega ,\,}Tr\mathsf{\,N}).  \tag{5}
\end{equation}
We could alternatively have used vectors $2\mathbf{n}=\mathsf{N\dot{\times}I}
$ and $2\mathbf{\Omega }=\mathsf{K\dot{\times}I.}$ The twelve first order
dyadic equations arising from the two additional HSOF $2-$forms then are 
\begin{equation}
\mathsf{(N\dot{\times}I-}\mathbf{a}\mathsf{)}\mathbf{\dot{ }}\;\mathsf{%
+\;\nabla (}Tr\mathsf{\,K)+(}Tr\mathsf{\,K)\,}\mathbf{a\;}\mathsf{+\;K\cdot
(N\dot{\times}I-}\mathbf{a}\mathsf{)+\omega \times (N\dot{\times}I-}\mathbf{a%
}\mathsf{)=\,}0  \tag{6}
\end{equation}
\begin{equation}
\mathsf{\nabla \times (N\dot{\times}I-}\mathbf{a}\mathsf{)+(}Tr\mathsf{%
\,K)\,K\dot{\times}I=}\;0  \tag{7}
\end{equation}
\begin{equation}
\mathsf{(K\dot{\times}I-\omega )}\mathbf{\dot{ }}-\mathsf{\nabla }\left( Tr%
\mathsf{\,N}\right) \;-\mathsf{\;}\left( Tr\mathsf{\,N}\right) \mathsf{\,}%
\mathbf{a\;}+\mathsf{\;K\cdot (K\dot{\times}I-\omega )+\omega \times (K\dot{%
\times}I)=\,}0  \tag{8}
\end{equation}
\begin{equation}
\mathsf{\nabla \times (K\dot{\times}I-\omega )-(}Tr\mathsf{\,N)\,K\dot{\times%
}I=\,}0.  \tag{9}
\end{equation}

The sets of first order equations we have derived, where the dyadic
components are taken as dependent variables and the four $\omega ^i$ form a
basis, may alternatively be analyzed by another of Cartan's techniques. 
\textbf{We must know that the set includes all integrability conditions},
and it must also be possible to write the exterior differential system such
that the left hand sides \textbf{linearly }involve exterior derivatives of
the dependent variables (e.g., no terms of the form $dK_{ab}\wedge dN_{cd}$
are allowed), while the right hand sides only involve forms in the adopted
basis (here $\omega ^i$, $\omega ^i\wedge \omega ^j,$ etc.). So-called 
\textbf{reduced} characters $s_i^{\prime }$ are then conveniently computed
from the left-hand sides alone, and \textbf{Cartan's test }is to calculate 
\[
h=\sum_{i=0}^{g-1}(g-i)\cdot s_i^{\prime }\,. 
\]
If $h$ is equal to the number of independent first order equations, this
establishes involutivity, i.e., the well set nature of the problem.
Moreover, if 
\[
s_g^{\prime }=n-g-\sum_{i=0}^{g-1}s_i^{\prime }=0, 
\]
the Cauchy-Kowalewsky solutions are causal, i.e., determined by data given
on a $g-1$ dimensional surface. In the present context $g=4$.

The 52 general vacuum dyadic equations were of this form \cite{E&W65} \cite
{E&U}. They result from computing the exterior derivatives of (2) and (3),
so have six $2-$forms and six $3-$forms in 34 dependent variables. Their
left hand sides are 
\[
dN_{ba}\wedge \omega ^b-d\overline{\omega }_a\wedge \omega ^4 
\]
\[
dK_{ba}\wedge \omega ^b+da_a\wedge \omega ^4 
\]
\[
dA_{ab}\wedge \omega ^b\wedge \omega ^4-\tfrac 16\,dB_{ab}\wedge \omega
^c\wedge \omega ^d\,\varepsilon _{.cd}^b 
\]
\[
dB_{ab}\wedge \omega ^b\wedge \omega ^4+\tfrac 16\,dA_{ab}\wedge \omega
^c\wedge \omega ^d\,\varepsilon _{.cd}^b\,. 
\]
We calculate $s^{\prime }=\left\{ 0,6,12,10\right\} ,$ so $h=52.$ There are
however $s_4^{\prime }=6$ arbitrary functions in the solution, and this
system is therefore not causal. The new HSOF system of dyadic equations
adjoins 12 equations in two additional $2-$forms to express equations
(6)--(9). Their left hand sides are 
\[
d(2n_a-a_a)\wedge \omega ^a+d(Tr\,\mathsf{K)\wedge \,}\omega ^4 
\]
\[
d(2\Omega _a-\overline{\omega }_a)\wedge \omega ^a-d(Tr\,\mathsf{N)\wedge \,}%
\omega ^4. 
\]
The reduced characters are $s^{\prime }=\{0,8,14,12\}$, $h=64$, $s_4^{\prime
}=0$, and therefore the final system is causal.

\section{The FOSH dyadic equations}

Linear combinations of the 52 general dyadic equations and the 12 new
conditions due to Nester can now easily be made to put the result in FOSH
form. The result is 34 equations involving the time derivatives and
symmetric space derivatives of the dyadic components and $\mathsf{A}$ and $%
\mathsf{B}$, and 30 ``constraint'' or transverse relations not involving
time derivatives. We give them in the following, written in full with their
quadratic right hand sides. For the connection components we obtain 
\begin{equation}
\mathbf{\dot{a}}\mathsf{-\nabla \cdot K}^{\QTR{sc}{T}}\mathsf{+\nabla \times
\omega =-K\dot{\times}N-\omega \cdot N+(}Tr\mathsf{\,N)\omega \,+\,}2\mathsf{%
K\cdot }\mathbf{n}  \tag{10}
\end{equation}
\begin{equation}
\mathbf{\dot{\QTR{mathsf}{\omega}}}\mathsf{+\nabla \cdot N}^{\QTR{sc}{T}}%
\mathsf{-\nabla \times \,}\mathbf{a\;}\mathsf{=-(}Tr\mathsf{\,N)}\mathbf{a\,}%
\mathsf{+\,(}2\mathbf{\Omega }\mathsf{-\omega )\cdot K-}2\mathsf{(}Tr\mathsf{%
\,K)}\mathbf{\Omega \,}\mathsf{-\,}2\mathbf{n}\mathsf{\cdot N}  \tag{11}
\end{equation}
\begin{equation}
\mathbf{\dot{\QTR{mathsf}{K}}}\,-\,\mathbf{a}\mathsf{\nabla \,-\,}2\mathsf{%
\nabla }\mathbf{n\,}\mathsf{+\,}2\mathbf{n}\mathsf{\nabla =-K\cdot K\,-\,}2%
\mathsf{(}Tr\mathsf{\,K)}\mathbf{\Omega }\mathsf{\times I-\omega \times
K+K\times \omega \,+\,}\mathbf{aa\,}\mathsf{-\,A}  \tag{12}
\end{equation}
\begin{equation}
\mathbf{\dot{\QTR{mathsf}{N}}}+\mathsf{\omega \nabla \,+\,}2\mathsf{\nabla }%
\mathbf{\Omega \,}\mathsf{-\,}2\mathbf{\Omega }\mathsf{\nabla =-K\cdot
N-\omega \times N+K\times }\mathbf{a\,}\mathsf{-\,}\mathbf{a}\mathsf{\omega
\,-\,}2\mathsf{(}Tr\mathsf{\,N)}\mathbf{\Omega }\mathsf{\times I+B.} 
\tag{13}
\end{equation}

We mention in passing that using $\mathbf{D}$ instead of $\mathsf{\nabla }$
usually makes the right hand sides of these equations less concise. The
prominent exception to this statement is the first equation, which becomes
homogeneous and linear, namely, 
\[
\mathbf{\dot{a}\,}\mathsf{-\,}\mathbf{D}\mathsf{\cdot K}^{\QTR{sc}{T}}\,%
\mathsf{+\,}\mathbf{D}\mathsf{\times \omega =\,}0. 
\]
It has been convenient to use $\mathbf{n}$ and $\mathbf{\Omega }$\textbf{, }%
and $\mathsf{\nabla }$ can operate from the right as well as from the left,
to express transposed indices. It is best to return to the nine components
of $\mathsf{K}$ and of $\mathsf{N}$ to see the FOSH structure; e.g. write $(2%
\mathsf{\nabla }\mathbf{n)}_{11}=\nabla _1N_{23}-\nabla _1N_{32},$ $(2%
\mathbf{n}\mathsf{\nabla }\mathbf{)}_{23}=\nabla _3N_{31}-\nabla _3N_{13},$ $%
(2\mathbf{\Omega }\mathsf{\nabla }\mathbf{)}_{22}=\nabla _2K_{31}-\nabla
_2K_{13},$ etc. The left hand sides of the FOSH equations (10)--(13) are
shown in Figure 1 as the product of a linear operator and a column vector.

The 24 constraint equations are 
\begin{equation}
\mathsf{\nabla \times K=\,}2\mathbf{\Omega a}-\mathsf{B}  \tag{14}
\end{equation}
\begin{equation}
\mathsf{\nabla \times N=-}\tfrac 12\mathsf{N}^{\QTR{sc}{T}} 
\begin{array}{c}
\times \\ 
\times
\end{array}
\mathsf{N\,-\,}\tfrac 12\mathsf{K} 
\begin{array}{c}
\times \\ 
\times
\end{array}
\mathsf{K+(}\mathbf{\Omega }\mathsf{\cdot K)\times I\,-\,}2\mathbf{\Omega }%
\mathsf{\omega -A}  \tag{15}
\end{equation}
\begin{equation}
\mathsf{\nabla \times (}\mathbf{a}\mathsf{-}2\mathbf{n}\mathsf{)=}2\mathsf{(}%
Tr\mathsf{\,K)}\mathbf{\Omega }  \tag{16}
\end{equation}
\begin{equation}
\mathsf{\nabla \times (\omega -}2\mathbf{\Omega }\mathsf{)=-}2\mathsf{(}Tr%
\mathsf{\,N)}\mathbf{\Omega }  \tag{17}
\end{equation}
By taking a second time derivative of the FOSH equations, interchanging
space and time derivatives, and substituting back both the FOSH and the
constraint equations, it can be seen that the dyadic variables all propagate
causally along null cones.

The 10 FOSH equations for the Weyl components are those for traceless
transverse massless spin$-2$ fields (dyadic and Bianchi identities, linearly
combined) and are as follows 
\[
2\,\mathbf{\dot{\QTR{mathsf}{B}}-\nabla \times A+A\times \nabla =} 
\]
\begin{equation}
2\mathsf{B\times \omega -2\omega \times B-2A\times }\mathbf{a\,}\mathsf{+\,}%
2\mathbf{a}\mathsf{\times A+K}^{\QTR{sc}{T}}\mathsf{\cdot B+B\cdot K\,-\,}2%
\mathsf{(}Tr\mathsf{\,K)B+K} 
\begin{array}{c}
\times \\ 
\times
\end{array}
\mathsf{B+B} 
\begin{array}{c}
\times \\ 
\times
\end{array}
\mathsf{K}  \tag{18}
\end{equation}
\[
2\,\mathbf{\dot{\QTR{mathsf}{A}}+\nabla \times B-B\times \nabla =} 
\]
\begin{equation}
2\mathsf{A\times \omega -2\omega \times A+2B\times }\mathbf{a\,}\mathsf{-\,}%
2\mathbf{a}\mathsf{\times B+K}^{\QTR{sc}{T}}\mathsf{\cdot A+A\cdot K\,-\,}2%
\mathsf{(}Tr\mathsf{\,K)A+K} 
\begin{array}{c}
\times \\ 
\times
\end{array}
\mathsf{A+A} 
\begin{array}{c}
\times \\ 
\times
\end{array}
\mathsf{K}  \tag{19}
\end{equation}
The left hand sides of the Bianchi equations are shown in Figure 2 as the
product of a linear operator and a column vector.

The final 6 constraint equations are 
\begin{equation}
\mathsf{\nabla \cdot A=-K\dot{\times}B\,-\,}4\mathbf{\Omega }\mathsf{\cdot B}
\tag{20}
\end{equation}
\begin{equation}
\mathsf{\nabla \cdot B=K\dot{\times}A\,+\,}4\mathbf{\Omega }\mathsf{\cdot A.}
\tag{21}
\end{equation}

Finally, it should be remarked that our derivations from well set exterior
differential systems obviate any need to verify that the transverse (or
constraint) equations, i.e., those not involving time derivatives, are
compatible with the FOSH system, and that they are propagated invariantly by
it.

\section{Appendix 1: Harmonic and co-moving coordinates and an explicit line
element}

The formulation of this paper is coordinate free and gauge independent. It
may however also be of use to briefly record how a harmonic time coordinate,
that is, one which satisfies a wave equation, and co-moving (with $_4\lambda
^\mu $) spacelike coordinates can be adopted.

In coordinate language put $t_{;\mu \nu }\,g^{\mu \nu }=0\;(\mu ,\nu
=1,\cdots ,4)$. Introducing fields $\phi =\mathbf{\dot{\QTR{mathit}{t}}},$ $%
\mathbf{A}=\mathsf{\nabla }t,$ the first order tetrad equations for this are 
\[
\mathbf{\dot{A}}-\mathsf{\nabla }\phi =-\mathsf{K}\cdot \mathbf{A}-\mathbf{%
\omega }\times \mathbf{A}+\phi \,\mathbf{a} 
\]
\[
\mathbf{\dot{\QTR{mathit}{\phi}}}-\mathsf{\nabla }\cdot \mathbf{A}=\mathbf{a}%
\cdot \mathbf{A}-\phi \,(Tr\,\mathsf{K}) 
\]
\[
\mathsf{\nabla }\times \mathbf{A}=2\,\phi \,\mathbf{\Omega .} 
\]
This set consists of four FOSH equations plus three constraints. It can be
added to, and solved simultaneously with, the equations in Section 4.

Co-moving spacelike coordinates, say $\mathbf{\dot{\QTR{mathit}{x}} }^\alpha
\;(\alpha =1,\cdots ,3),$ such that $\dot{x}^\alpha =0$ are found by setting 
\[
\mathbf{e}^\alpha =\,\mathsf{\nabla }x^\alpha , 
\]
and the integrability conditions for this are 
\[
\mathbf{\dot{e}}^\alpha =-\mathsf{K}\cdot \mathbf{e}^\alpha -\mathsf{\omega }%
\times \mathbf{e}^\alpha 
\]
\[
\mathsf{\nabla }\times \mathbf{e}^\alpha =0. 
\]
The $\mathbf{e}^\alpha $ are, understandably, the first dependent variables
we have found whose causal propagation is strictly timelike and not along
the local null cone. We introduce coordinate components $\mathbf{A}\cdot 
\mathbf{e}^\alpha =A^\alpha $ and $h^{\alpha \beta }=\mathbf{e}^\alpha \cdot 
\mathbf{e}^\beta $, and also calculate the inverse $h_{\alpha \beta }$ $%
(h^{\alpha \gamma }\,h_{\alpha \beta }=\delta _\beta ^\gamma ).$ Then $\phi
, $ $A_\alpha =h_{\alpha \beta }A^\beta $ and $h_{\alpha \beta },$ functions
of $x^\alpha ,$ and $t,$ enter the final line element: 
\[
ds^2=-\phi ^{-2}\,dt^2+2\phi ^{-2}A_\alpha \,dx^\alpha dt+(h_{\alpha \beta
}-\phi ^{-2}A_\alpha \,A_\beta )\,dx^\alpha \,dx^\beta . 
\]
In covariant $4-$vector terms, the coordinate components of the tetrad
vectors take the form 
\[
_r\lambda ^\mu =\left( 
\begin{array}{cc}
_a\lambda ^\alpha & _aA \\ 
0 & \phi
\end{array}
\right) 
\]
\[
^r\lambda _\mu =\left( 
\begin{array}{cc}
^a\lambda _\alpha & 0 \\ 
-\phi ^{-1}A_\alpha & \phi ^{-1}
\end{array}
\right) 
\]
The orthonormal triad components of the vector $\mathbf{A}$ are written $%
_a\lambda ^\alpha \,A_\alpha .$

\section{Appendix 2: $2+1$ dimensional gravity}

An entirely parallel development can be made using special orthonormal
frames (SOF) in $2+1$ gravity. The exterior differential system for
immersion of $3-$dimensional flat spaces in the $21$ dimensional orthonormal
frame bundle over $6$ dimensional flat space, which is generated by the
usual immersion forms $\omega ^A$ and $d\omega ^A\,(A=4,5,6),$ and the
Riemann $2-$forms $R_b^a\,(a,b,c=1,2,3),$ has $s=\left\{ 3,6,3\right\} $ and 
$g=9.$ We add a $2-$form 
\[
\omega _a^b\wedge \omega ^c\varepsilon _{\cdot bc}^a+y_a\omega ^b\wedge
\omega ^c\varepsilon _{\cdot bc}^a, 
\]
a $3-$form 
\[
\omega _{ab}\wedge \omega ^a\wedge \omega ^b-z\omega ^1\wedge \omega
^2\wedge \omega ^3, 
\]
and their closures, to introduce 4 new variables $y_a$ and $z$, so the
exterior differential system is prolonged to a total of $25$ dimensions. The
SOF conditions are imposed as two additional closed exterior forms, namely, 
\[
dz 
\]
\[
d(y_a\omega ^a). 
\]
These adjoin solutions of a linear Dirac equation \cite{N89}. Now the Cartan
characters show the exterior differential system to be well set and causal
with $s=\left\{ 4,8,7\right\} $, and $g=6.$ Solutions are $3-$parameter
co-frame rotation bundles over $3-$space, determined by causal integration
from $2-$spaces.

Orthonormal triad formalism for $3-$dimensional geometry yields $9$
equations for $9$ Ricci rotation coefficients grouped in a dyadic $\mathsf{N}
$: 
\[
\mathsf{\nabla }\times \mathsf{N\,+\,}\tfrac 12\mathsf{N}^{\QTR{sc}{T}} 
\begin{array}{c}
\times \\ 
\times
\end{array}
\mathsf{N=\,}0. 
\]

The SOF\ conditions require the identification of $z$ as $Tr\mathsf{\,N}$%
\textsf{\ }and the $y_a$ as the components of $\mathbf{n}=\tfrac 12\mathsf{N%
\dot{\times}I.}$ To the nine equations for a flat $3-$space are added six
SOF equations: 
\[
\mathsf{\nabla (}Tr\mathsf{\,N)=\,}0 
\]
\[
\mathsf{\nabla \times \,}\mathbf{n\,}\mathsf{=\,}0. 
\]
Inserting a $-1$ corresponding to timelike nature of the $3$ direction, the
complete set falls into FOSH form with $6$ transverse equations. The left
hand sides of the FOSH\ equations are shown in Figure 3 as the product of a
linear operator and a column vector.

\section{Acknowledgments}

FBE acknowledges with thanks a helpful extended conversation with Jim Nester
at the Seventh Marcel Grossmann meeting in 1994. RSR acknowledges the
enduring influence and inspiration of the late Hanno Rund. This research was
supported by the National Aeronautics and Space Administration.

\end{document}